\documentclass{emulateapj} 

\usepackage{epsfig}

\usepackage{url}
\usepackage{graphics,graphicx}
\usepackage{subfigure}
\usepackage{epsfig}
\usepackage{lscape}
\usepackage{longtable}
\usepackage{url}
\usepackage{bigstrut,bigdelim,multirow}
\usepackage{multirow}

\begin{document}
\shorttitle{Double Burst GRB 110709B}
\shortauthors{Zhang et al.}
\title{Unusual Central Engine Activity in the Double Burst GRB 110709B}

\author{Bin-Bin Zhang\altaffilmark{1}, David N. Burrows\altaffilmark{1}, Bing Zhang\altaffilmark{2}, Peter {M\'esz\'aros}\altaffilmark{1,3}, Xiang-Yu Wang\altaffilmark{4,5}, Giulia Stratta\altaffilmark{6,7}, Valerio D'Elia\altaffilmark{6,7}, Dmitry Frederiks\altaffilmark{8}, Sergey Golenetskii\altaffilmark{8}, Jay R. Cummings\altaffilmark{9,10}, Jay P. Norris\altaffilmark{11}, Abraham D. Falcone\altaffilmark{1}, Scott D. Barthelmy\altaffilmark{12}, Neil Gehrels\altaffilmark{12}}

\altaffiltext{1}{Department of
Astronomy and Astrophysics, Pennsylvania State University,
University Park, PA 16802, USA; bbzhang@psu.edu} 
\altaffiltext{2}{Department of
Physics, University of Nevada, Las Vegas, NV 89154, USA}

\altaffiltext{3}{Department of Physics, Pennsylvania State
University, University Park, PA 16802, USA}

\altaffiltext{4}{Department of Astronomy, Nanjing University, Nanjing, 210093, China}

\altaffiltext{5}{Key laboratory of Modern Astronomy and Astrophysics (Nanjing University), Ministry of Education, Nanjing 210093, China}

\altaffiltext{6}{ASI-Science Data Center, Via Galileo Galilei, I-00044 Frascati, Italy}

\altaffiltext{7}{INAF-Osservatorio Astronomico di Roma, Via Frascati 33, I-00040 Monteporzio Catone, Italy}

\altaffiltext{8}{Ioffe Physico-Technical Institute, Laboratory for Experimental Astrophysics, 26 Polytekhnicheskaya, St Petersburg 194021, Russian Federation}

\altaffiltext{9}{CRESST and NASA/GSFC, Greenbelt, MD 20771, USA}

\altaffiltext{10}{University of Maryland, Baltimore County, 1000 Hilltop Circle, Baltimore, MD 21250, USA}

\altaffiltext{11}{Physics Department, Boise State University, 1910 University Drive, Boise, ID 83725, USA}

\altaffiltext{12}{NASA Goddard Space Flight Center, Greenbelt, MD 20771, USA}
\begin{abstract}

The double burst, GRB 110709B, triggered {\em Swift}/BAT twice at 21:32:39 UT and 21:43:45 UT, respectively, on 9 July 2011. This is the first time we observed a GRB with two BAT triggers. In this paper, we present simultaneous {\em Swift} and Konus-{\em{WIND}} observations of this unusual GRB and its afterglow. If the two events originated from the same physical progenitor, their different time-dependent spectral evolution suggests they must belong to different episodes of the central engine, which may be a magnetar-to-BH accretion system.

\end{abstract}


\section{Introduction}

Gamma-Ray Bursts (GRBs) have been thought to be non-repeatable events through both observation and theoretical understanding. The general picture of a GRB is as follows: (1) A ``central engine" consisting of a rapidly rotating black hole (BH) and a nuclear-density accretion disk is formed from a progenitor system, which invokes either core-collapse of a massive star (Woosley 1993; MacFadyen \& Woosley 1999; Fryer et al. 2007) or merger of two compact stellar objects such as NS-NS or BH-NS \citep{paczynski86,eichler89,paczynski91,narayan92}. (2) Relativistically expanding ejecta composed of many mini-shells with a wide-range of Lorentz factors are launched by the central engine. Internal shocks \citep{rees94} are formed during the collisions of those shells and produce the observed prompt GRB emission (mostly in Gamma-ray band). Observationally this is the phase when GRBs trigger gamma-ray band detectors. (3) The ejecta are further decelerated by an ambient medium (e.g., interstellar medium; ISM) and produce a long-term broadband afterglow through an external-forward shock (M\'esz\'aros \& Rees 1997; Sari et al. 1998) and/or external-reverse shock (M\'esz\'aros \& Rees 1997, 1999; Sari \& Piran 1999a,b). (4) In some cases, the central engine can be restarted during the afterglow phase and X-ray flares are produced through dissipation of a late wind launched from a long-lasting central engine (Burrows et al. 2005a; Zhang et al. 2006; Fan \& Wei 2005; Ioka et al. 2005; Wu et al. 2005; Falcone et al. 2006; Romano et al. 2006a; Lazzati \& Perna 2007; Maxham \& Zhang 2009; see Zhang 2007 for review). Although X-ray flares are generally regarded to arise from the same physical region as prompt emission, they release their energy mostly in the soft X-ray band.

GRB 110709B triggered the Burst Alert Telescope on-board {\em Swift} (Gehrels et al. 2004) twice. Each of the triggers, separated by 11 minutes, consists of an otherwise typical long GRB light curve in the hard X-ray/gamma-ray band. X-ray observations during the second burst show that this event also produced bright soft X-ray emission. This provides a rare opportunity to conduct a detailed broadband study of the central engine properties. 

In this paper, we first report the {\em Swift} and Konus-{\em{WIND}} observations of GRB 110709B in \S 2. Then we present multi-wavelength spectroscopy and timing studies in \S 3. The physical implications on the central engine properties are discussed in \S 4. We draw our conclusions in \S 5.

\section{Observations and Data Analysis}

\subsection{{Swift} Data}

GRB 110709B first triggered the Burst Alert Telescope (BAT; Barthlmy et al. 2005) on-board {\em Swift} at 21:32:39 UT on 9 July 2011 (Cummings et al. 2011a). {\em Swift} slewed immediately to the burst. The two narrow field instruments, the X-ray Telescope (XRT; Burrows et al. 2005b) and the Ultraviolet Optical telescope (UVOT; Roming et al. 2005) on-board {\em{Swift}} began to observe the field at $T_0+80.5$ seconds and $T_0+91 $ seconds, respectively, where $T_0$ is the BAT trigger time. A bright X-ray afterglow was localized at ${\rm R.A. (J2000)}=10^h58^m37.08^s$, ${\rm Dec. (J2000)}=-23^{\circ}27'17\farcs 6$ with an uncertainty of 1\farcs4 (90\% confidence, Beardmore et al. 2011). No reliable optical source was found within the XRT error circle (Holland et al. 2011a,b).

Interestingly, at 21:43:25 UT on 9 July 2011, 11 minutes after the first trigger, the BAT was triggered again and located a second event from the same location (Barthelmy et al. 2011). The second outburst has comparable intensity and light curve characteristics to the first outburst. Regarding the two outbursts as two episodes of a single burst, the separation (11 minutes) is the longest compared to other multi-episode GRBs measured by {\em Swift}. In this paper, we use the term ``double burst" to stress the unusual nature of this double-trigger GRB. We will use the term ``the first sub-burst" to refer to the first outburst and ``the second sub-burst" to refer to the second outburst. However, as we will show below, the two events are clearly related, indicating that they originated from the same physical progenitor system (see Drago \& Pagliara 2007 for a statistical study of similar GRBs with long quiescent phases).

\begin{figure}

\begin{center}
\resizebox{3.6in}{!}{\includegraphics{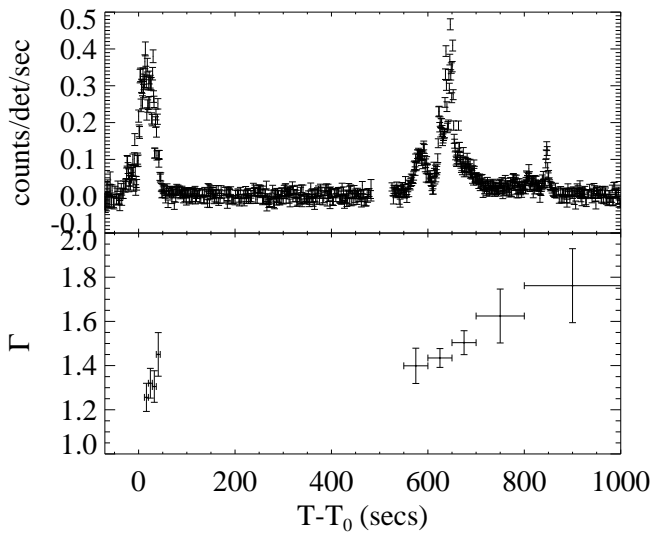}}
\end{center}

\caption{BAT count rates (upper panel) and photon index evolution (lower panel) of GRB 110709B. The spectral model is a simple power law.}
\label{fig:batlc}
\end{figure}

We processed the {\em Swift}/BAT data using standard {HEAsoft} tools (version 6.11). As shown in Fig. \ref{fig:batlc}, the first sub-burst lasted from $T_0-28$ seconds to $T_0+55$ seconds with $T_{90,1st}=55.6\pm 3.2$ seconds. The second sub-burst lasted from $\sim T_0+550$ seconds to about $T_0+865$ seconds with $T_{90,2nd}=259.2\pm 8.8$ seconds (Cummings et al. 2011b). There was no flux detectable in BAT from about $T_0+180$ seconds to about $T_0+550$ seconds. We extracted the BAT spectra in several slices. The lower panel in Fig. \ref{fig:batlc} shows the photon indices obtained by fitting the spectra with a simple power law model. It is obvious that both sub-bursts have strong hard-to-soft spectral evolution. The photon indices range from $\sim 1.25$ to $\sim 1.75$. The BAT band (15-150 keV) fluences of the first and second sub-bursts are $8.95_{-0.62}^{+0.29} \times 10^{-6}\ {\rm erg} \ {\rm cm}^{-2} $ and $1.34_{-0.07}^{+0.05} \times 10^{-5}\ {\rm erg} \ {\rm cm}^{-2}$ respectively.
 
\begin{figure}
\begin{center}
\resizebox{3.4in}{!}{\includegraphics{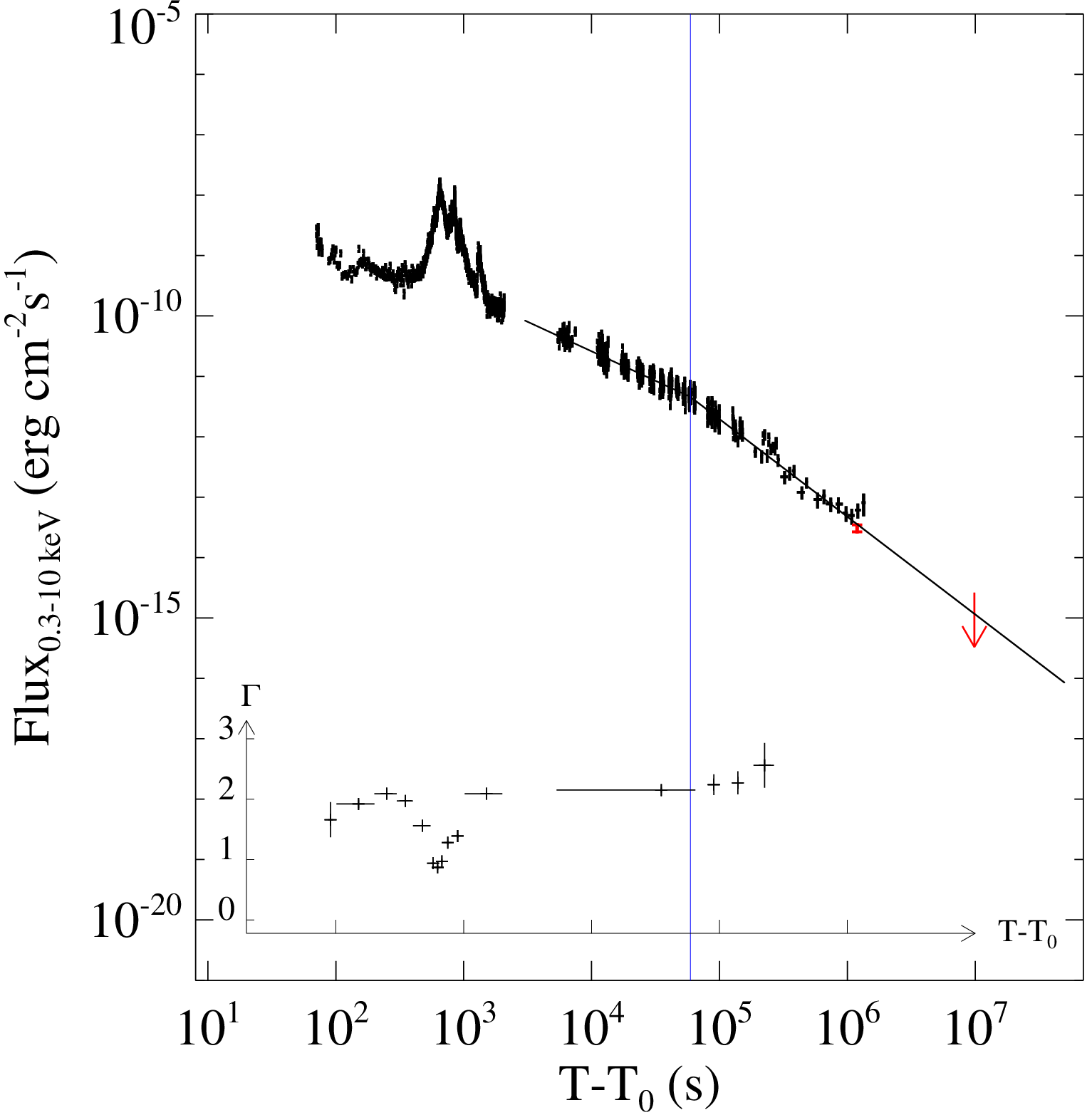}}
\end{center}
\caption{{\em Swift}/XRT light curve of GRB 110709B. The inner plot shows the photon index evolution. Red points are from Chandra observations (see \S \ref{sec:dark}). The black solid line shows the broken power-law fit to the lightcurve after the flare. The vertical blue solid line marks the break time, $t_b$ (see \S 2.1 for details).}

\label{fig:xrtlc}
\end{figure}

We processed the {\em Swift}/XRT data using our own IDL codes which employ the standard HEAsoft analysis tools. For technical details please refer to Zhang et al. 2007. Fig. \ref{fig:xrtlc} shows the XRT light curve and spectral evolution. The prolonged and energetic flaring activity continues up to $T_0+2000$ seconds, which corresponds to the second sub-burst time period. The light curve after the flare can be fitted by a broken power-law with $\alpha_1=0.98\pm0.08$, $\alpha_2=1.6\pm 0.13 $ and a break time $t_b=5.9\pm 4.1 \times 10^4\ s$ ($\chi^2/dof=338.6/279$). Assuming GRB 110709B is at the average redshift (z$\sim$2) of {\em Swift} GRBs, its rest frame break time, $t_{b,rest}$ (=$\frac{t_b}{1+z}\sim 2.0 \times 10^4 s$), and the corresponding X-ray luminosity, $L_{X,b}$ ($\sim 5\times 10^{46}erg \ s^{-1}$), are consistent with of the $t_{b,rest}$-$L_{x,b}$ correlation of previous {\em Swift} GRBs (Dainotti et al. 2010). The X-ray spectrum can be fitted with an absorbed power-law with total column ${\rm N_H}=2.14_{-0.21}^{+0.22} \times 10^{21} {\rm cm}^{-2}$, which includes the Galactic foreground ${\rm N_H}=5.6\times 10^{20} {\rm cm}^{-2}$ (D'Elia et al. 2011). Strong spectral evolution was observed in the second sub-burst phase, where the photon indices vary significantly from $\Gamma \sim 0.9$ to $\Gamma \sim 2.6$. The spectrum after $T_0+4000s$ has no significant evolution, with an average photon index $\Gamma \sim 2.1$. The total fluence in XRT band (0.3-10 keV) is $4.07\pm0.56 \times 10^{-6} \ {\rm erg} \ {\rm cm}^{-2}$. 

In order to check whether the break in the XRT light curve is due to curvature caused by an incorrect reference time $T_0$ effect (e.g., Yamazaki 2009 and Liang et al. 2009, 2010), we plotted the XRT light curve in reference to the trigger time of the second sub-burst. We found that the $t_b$, $\alpha_1$ and $\alpha_2$ do not significantly change within 1-sigma range. We thus conclude that the break is intrinsic.

\subsection{{\rm Konus}-{{Wind}} Data}

\begin{figure*}

\begin{center}
\resizebox{6.5in}{!}{\includegraphics{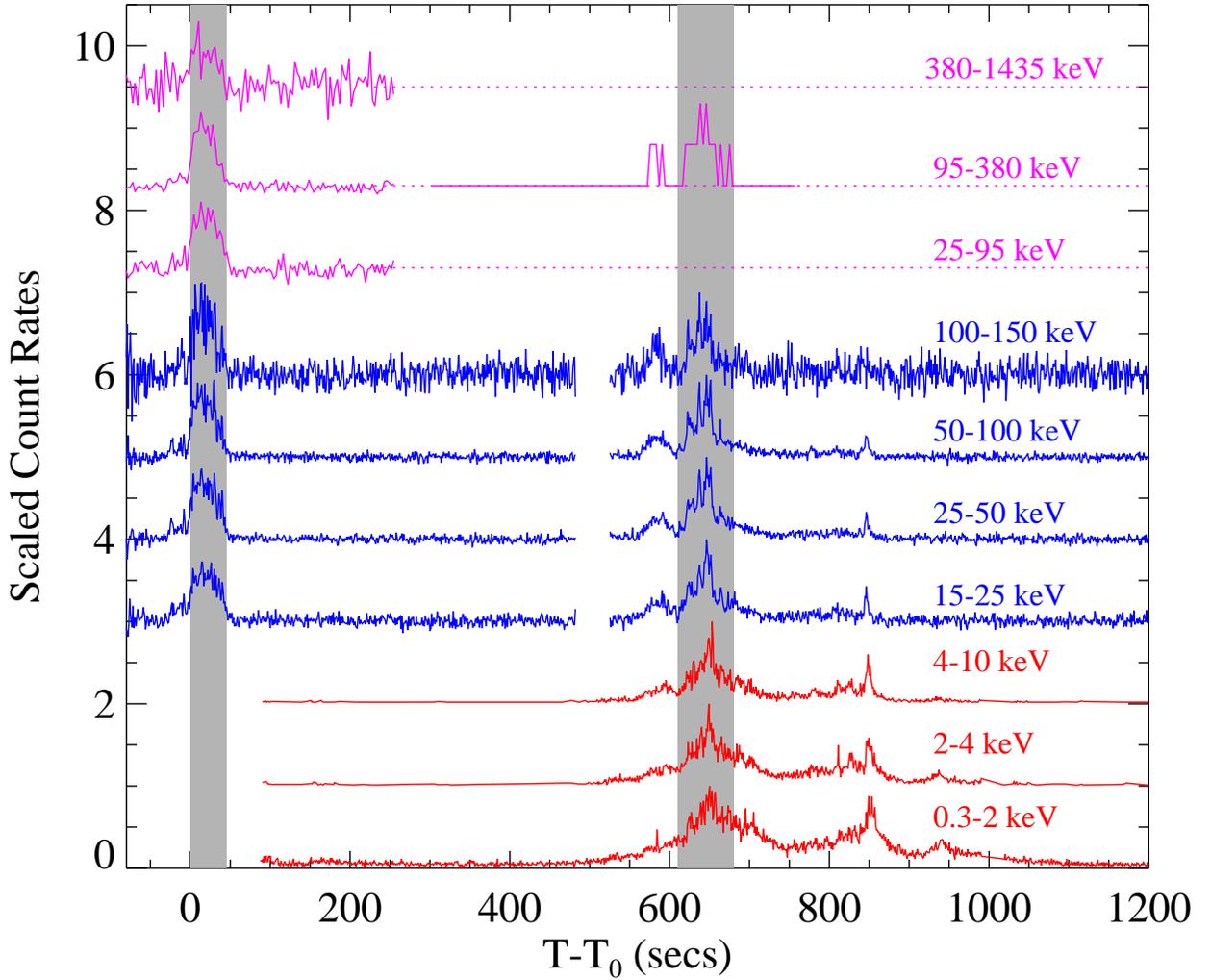}}
\end{center}

\caption{Multi-wavelength light curve for the prompt emission phase of GRB 110709B. Different colors indicate different instruments as follows: {\it Magenta}: Konus-{\em WIND}; {\it blue}: {\em Swift}/BAT; {\it red}: {\em Swift}/XRT. The shaded areas indicate Episodes I \& II that are selected to calculate spectral lags (see \S \ref{sec:lag} and Table 2 for more details). The pulse width evolution with energy, namely the pulses in softer bands tend to be broader, are similar with other GRBs (e.g., Romano et al. 2006b).}

\label{fig:multilc}
\end{figure*}

GRB 110709B triggered detector S1 of the Konus-{\em{WIND}} gamma-ray spectrometer (Apterkar et al. 1995) at 21:32:44.567 s UT on 9 July 2011 (Golenetskii et al. 2011). Konus-{\em{WIND}} recorded the first sub-burst with high-resolution data. The $T_{90}$ of the first sub-burst in Konus-{\em WIND} energy band (20 keV - 5 MeV) is $51.3\pm 7.6$ s. The fluence in the same energy range is $2.6\pm 0.2 \times 10^{-5}\ {\rm erg} \ {\rm cm}^{-2} $. The second sub-burst fell into a telemetry gap but was recorded by the instrument's spare count rate measurement channel (Fig. \ref{fig:multilc}). The overlap detection of the first sub-burst allows a BAT+Konus-{\em WIND} multi-wavelength study.

\section{Multi-wavelength timing and spectroscopy properties}

\subsection{Joint Spectral Fit}

\begin{figure}

\begin{center}
\resizebox{3.6in}{!}{\includegraphics[angle=270,scale=0.34]{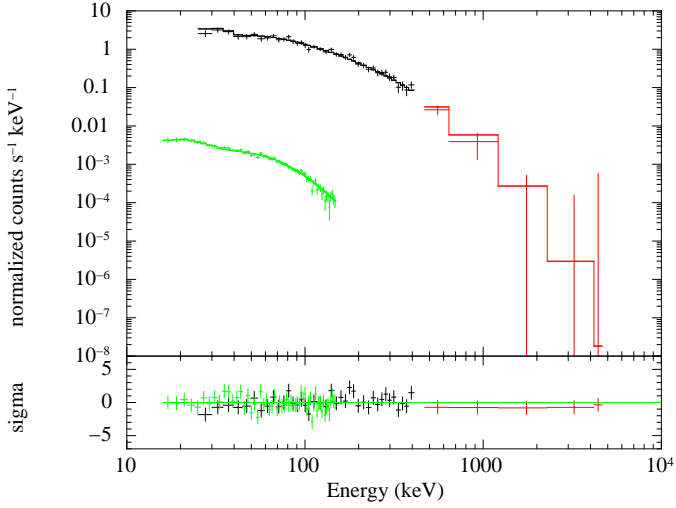}}
\end{center}

\caption{Joint fitting to the time-averaged Konus-{\em WIND}+{\em Swift}/BAT spectra between 3.594-44.810 s. {\it Green}: {\em Swift}/BAT spectrum. {\it Black} and {\it red}: Konus-{\em{WIND}} spectra. Solid lines are the best-fit model. Lower panel plots the residuals.}
\label{fig:batkwfit}
\end{figure}

\begin{figure}
\begin{center}
\resizebox{3.6in}{!}{\includegraphics[angle=270,scale=0.34]{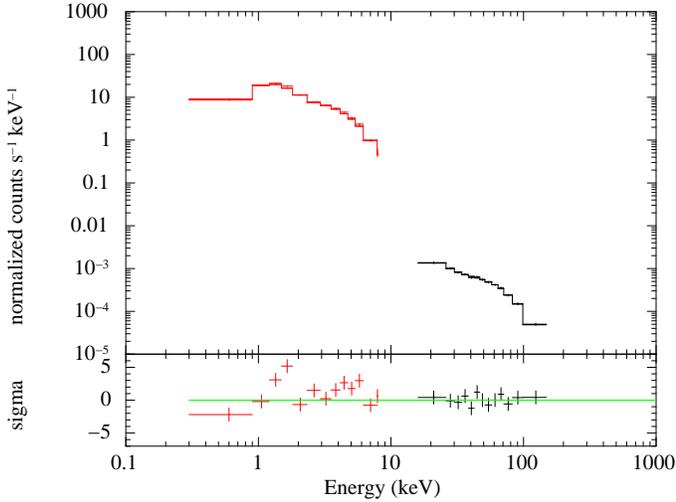}}
\end{center}

\caption{Joint fitting to the time-averaged {\em Swift}/BAT+{\em Swift}/XRT spectra between 550-1000 s. {\it Red}: {\em Swift}/XRT spectrum. {\it Black}: {\em Swift}/BAT spectrum. Solid lines are the best-fit model. Lower panel plots the residuals in terms of sigmas.}
\label{fig:batxrtfit}
\end{figure}

As shown in Fig \ref{fig:multilc}, the first sub-burst was simultaneously observed by Konus-{\em{WIND}} and {\em{Swift}}/BAT, so we are able to perform joint spectral fitting using the spectra of those two instruments. We divide the time period of the first sub-burst into 5 time slices. The exact time ranges of each slice are listed in Table 1. For the first four slices, the best fit model is a cut-off power-law (CPL, cutoffpl in Xspec 12). For the 5th slice, the best fit model is a simple power-law (PL, powerlaw in Xspec 12). The time-dependent fitting results are presented in Table 1. The time-integrated spectrum (3.594 s to 44.810 s ) can also be fitted with a cut-off power-law model with $\alpha=1.17\pm0.04$, $E_p=311_{-38}^{+45}$ and $\chi^2/{\rm dof}=125/129$ (Fig. \ref{fig:batkwfit}). The second sub-burst was simultaneously observed by {\em Swift}/BAT and {\em Swift}/XRT. Similarly with the first sub-burst, we are able to perform joint spectral fitting using the spectra of those two instruments. We divide the time period of the second sub-burst into 5 slices (listed in Table 1). We fit the spectrum of each slice using absorbed cut-off power-law model. An underlying simple power-law decaying component was also taken into account and subtracted using the same strategy as in Falcone et al. 2007. The time-dependent fitting results are presented in Table 1. The time-averaged (550s to 1000 s) BAT+XRT spectra are well fitted by the absorbed cut-off power-law model with $\alpha=1.12\pm0.04$, $E_p=116_{-8}^{+9}$ and $\chi^2/{\rm dof}=687/679$ (Fig. \ref{fig:batxrtfit}).

\begin{figure}

\begin{center}
\resizebox{3.1in}{!}{\includegraphics{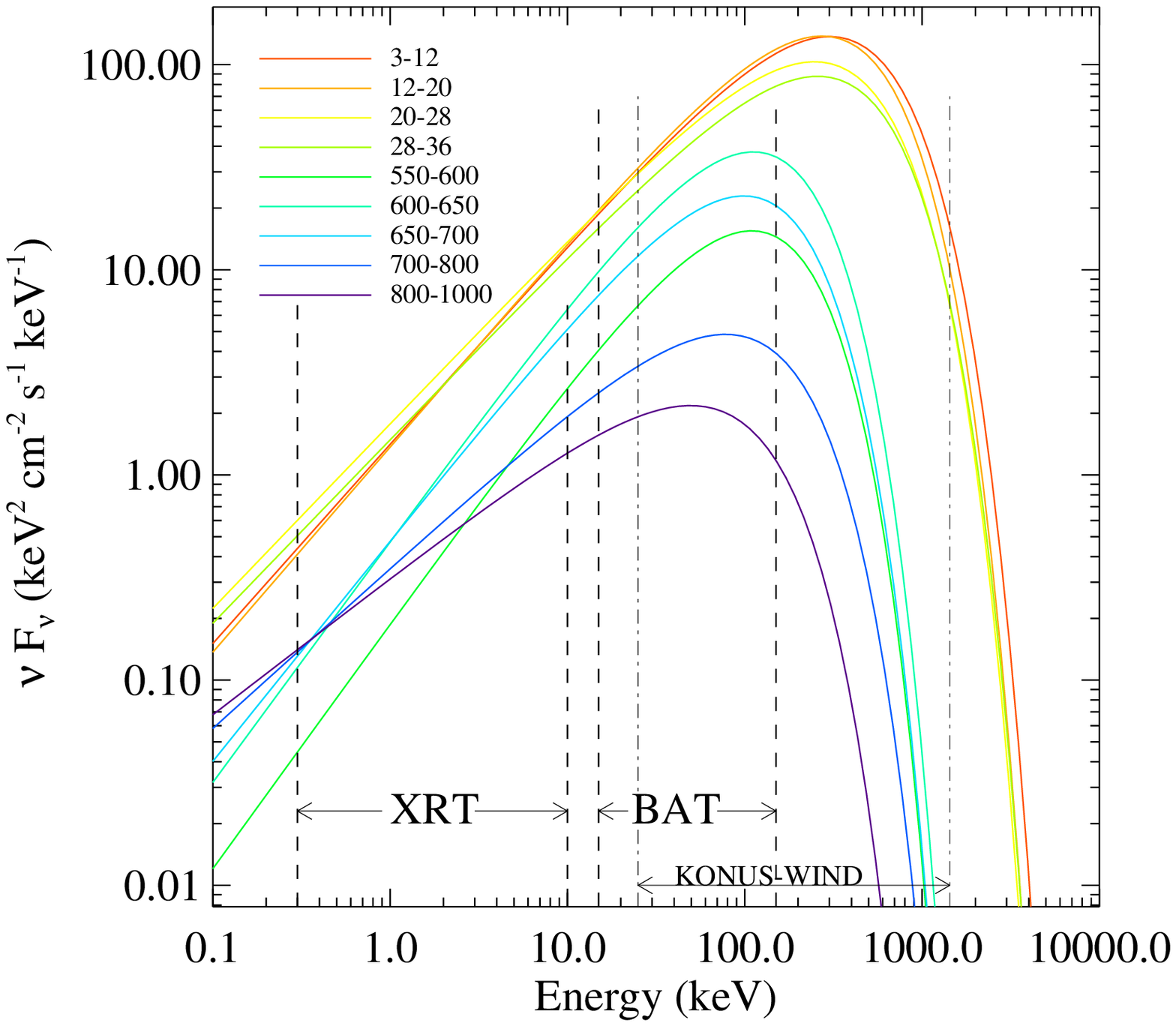}}
\end{center}
\caption{Modeled spectral energy distribution in different time intervals of the whole double burst period. Time intervals in seconds after $T_0$ are given in the legend.}
\label{fig:sed}
\end{figure}

\begin{figure}

\begin{center}
\resizebox{3.4in}{!}{\includegraphics{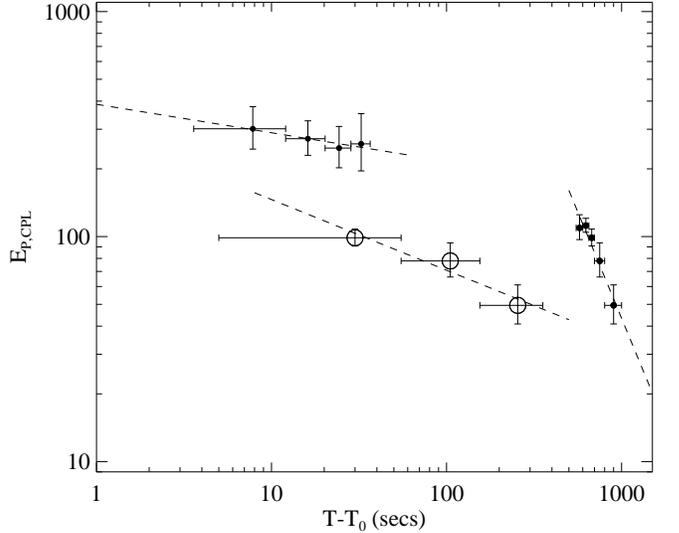}}
\end{center}
\caption{$E_p$ as a function of time. Dashed lines indicate the simple power-law fit. For the first sub-burst (filled circles), $E_p\propto t^{-0.13}$ while for the second sub-burst (filled circles), $E_p \propto t^{-1.9}$. Open circles show the $E_p$ evolution of the second sub-burst if $T_0$ is shifted to the trigger time of the second sub-burst, in which case $E_p\propto t^{-0.31}$ . }
\label{fig:ept}
\end{figure}

\begin{figure}

\begin{center}
\resizebox{3.4in}{!}{\includegraphics{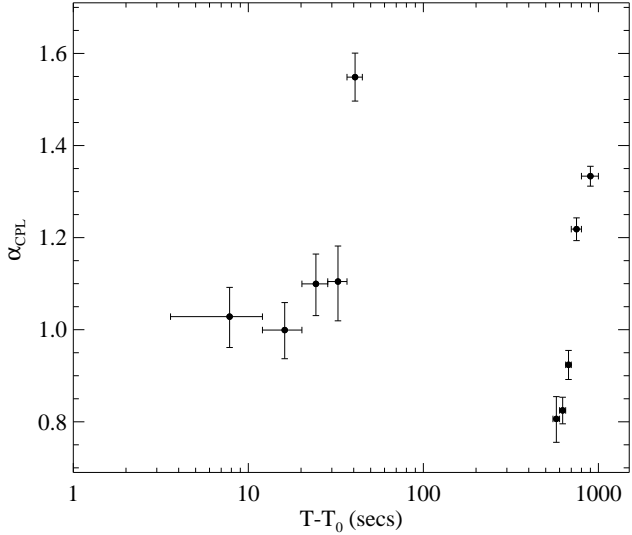}}
\end{center}

\caption{The spectral index of the cutoff power law model, $\alpha$, as a function of time.}
\label{fig:alphat}
\end{figure}

The spectral evolution during the whole double burst shows an overall hard-to-soft trend. In Fig. \ref{fig:sed}, we plot the modeled spectral energy distribution in different time intervals, which demonstrates the intrinsic spectral shape evolution. Fig. \ref{fig:ept} \& \ref{fig:alphat} show the evolution of $E_p$ and $\alpha$ respectively. Although strong spectral evolution is exhibited by both sub-bursts, their time-dependent behaviors are very different. For example, as shown in Fig \ref{fig:ept}, the $E_p$ of the first sub-burst decays to $\propto t^{-0.13}$ while the $E_p$ of the second sub-burst decays to $\propto t^{-1.9}$ (or $\propto t^{-0.31}$ if we shift reference time of the second sub-burst to its trigger time). The different time-dependent spectral of the two sub-bursts may suggest that the two sub-bursts are from different stages of the same central engine (see \S \ref{sec:theory} for more discussions).

\begin{table}

 
 \label{table1}
 
 \caption{Joint Fit Results} 
 \begin{tabular}{llllll}
 \hline
 \hline

\multicolumn{1}{c}{Time interval}   &   \multicolumn{1}{c}{Model }   &  \multicolumn{1}{c}{$\alpha$ } & \multicolumn{1}{c}{$E_p$ }   &  \multicolumn{1}{c}{$\chi^2/dof$}  & \multicolumn{1}{c}{Inst.} \\
 \multicolumn{1}{c}{s}      &   &  &  \multicolumn{1}{c}{keV} &  &   \\
\hline


(3.594,12.042) & CPL & $1.03\tablenotemark{a}\pm0.06$  &$301_{-57}^{+77}$  & 127/128 & BAT+KW\\

(12.042,20.230) & CPL & $1.0\pm0.06$  &$272_{-41}^{+53}$  & 135/128 & BAT+KW\\

(20.230,28.426) & CPL & $1.1_{-0.06}^{+0.07}$  &$247_{-46}^{+60}$  & 156/128 & BAT+KW\\

(28.426,36.618) & CPL & $1.1\pm0.08$  &$258_{-63}^{+94}$  & 111/128&BAT+KW\\

(36.618,44.810) & PL \ & $1.55\pm0.05$  &$-$  & 132/129&BAT+KW\\

(3.594,44.810) & CPL & $1.17\pm0.04$  &$311_{-38}^{+45}$   & 125/129&BAT+KW\\

(550,600) & CPL & $0.80\pm0.05$  &$109_{-12}^{+15}$   & 263/303&BAT+XRT\\
(600,650) & CPL & $0.82\pm0.03$  &$112_{-7}^{+9}$   & 360/418&BAT+XRT\\
(650,700) & CPL & $0.92\pm0.03$  &$99_{-8}^{+9}$   & 343/365&BAT+XRT\\
(700,800) & CPL & $1.22\pm0.02$  &$78_{-12}^{+16}$   & 438/456&BAT+XRT\\
(800,1000) & CPL & $1.33\pm0.02$  &$72_{-13}^{+17}$   & 512/501&BAT+XRT\\
(550,1000) & CPL & $1.12\pm0.01$  &$116_{-8}^{+9}$   & 687/679&BAT+XRT\\
\hline \hline
\end{tabular}

\footnotetext{$^a$ Low energy photon index $\alpha$ is defined by $C(E)\propto E^{-\alpha}{\rm exp}(-\frac{E(2+\alpha)}{E_p})$ for CPL and $C(E)\propto E^{-\alpha}$ for PL. Errors are given at the 1-sigma level.}

\end{table}

\subsection{$E_p$-$E_{\gamma,iso}$ Relation and Implication for Redshift}
\begin{figure}

\begin{center}
\resizebox{3.4in}{!}{\includegraphics{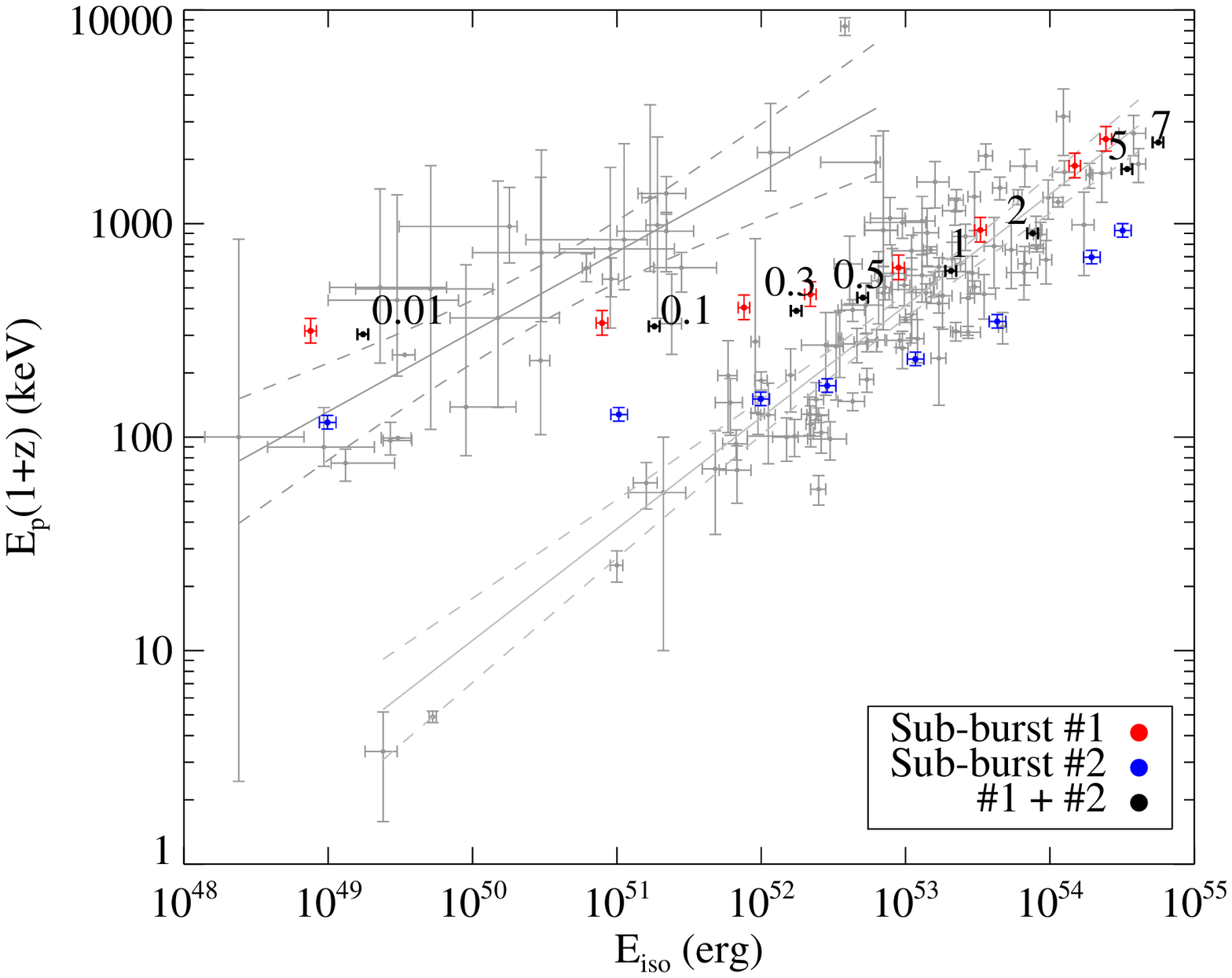}}
\end{center}

\caption{$E_p-E_{iso}$ plot made by putting the double burst at different assumed redshift $z_x$. {\it Red} : the first sub-burst. {\it Blue}: the second sub-burst. {\it Black}: the whole double burst. Background grey data points are the type I/short and type II/long samples taken from Zhang et al. 2009. Background solid lines indicate the best-fit $E_p-E_{\gamma,iso}$ correlation for both samples. The two dashed lines mark the best fits at 1 $\sigma$ confidence level. For each group of three points (red, blue and black), the nearest number indicates the corresponding assumed redshift $z_x$.}
\label{fig:epeiso}
\end{figure}

There has been no redshift measurement for GRB 110709B, so the rest-frame peak energy, $E_p(1+z)$, and the isotropic energy, $E_{\gamma,iso}$, are unknown. On the other hand, one can assume it has a redshift $z_{x}$ and plot the corresponding $E_p(z_x)$ and $E_{\gamma,iso}(z_x)$ on the $E_p-E_{\gamma,iso}$ (Amati relation; Amati et al. 2002) diagram. The well-known Amati relation suggests that most long (or type II; Zhang et al. 2009) bursts follow the $E_p \propto E_{\gamma,iso}^{1/2}$ track (Amati et al. 2002, Zhang et al. 2009). Since GRB 110709B is obviously a long burst (especially with two long sub-bursts), in principle it should fall into the same track as other typical long (type II) bursts. In Fig. \ref{fig:epeiso}, we assign GRB 110709B onto the $E_p$-$E_{\gamma,iso}$ diagram by assuming its redshift is in the range of $z_x=0.01-7$. 
It is interesting to note that, at the average redshift ($z\sim 2$) of {\em Swift} GRBs, GRB 110709B is well consistent with the previous Amati relation. 


\subsection{Spectral Lag}\label{sec:lag}

\begin{figure}

\begin{center}
\resizebox{3.4in}{!}{\includegraphics{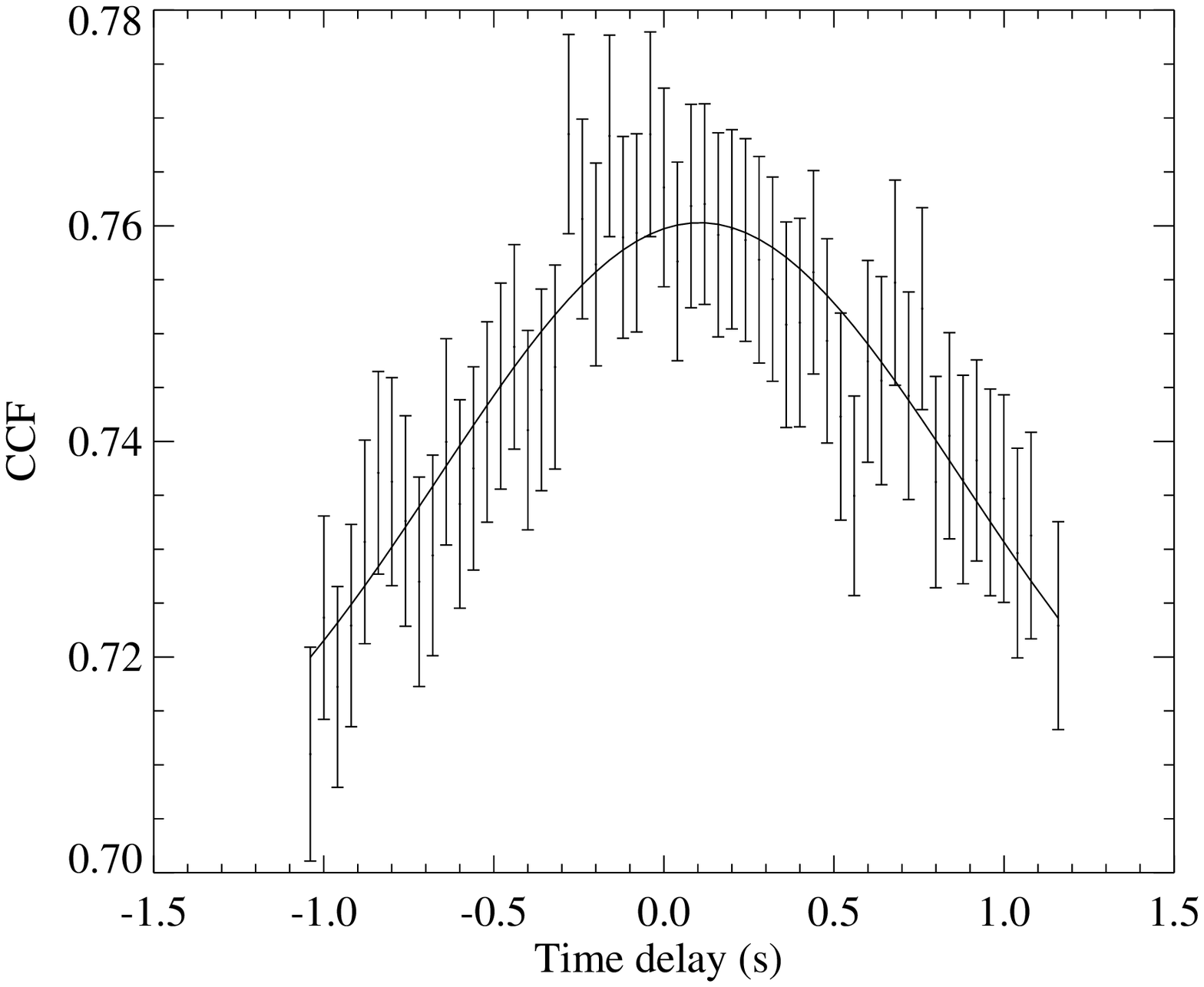}}
\end{center}

\caption{Cross-correlation function (CCF) vs time delay between the 25-50 keV and 50-100 keV channels for the second sub-burst in Episode II (see Fig. \ref{fig:multilc}) . The error in each bin of the CCF is determined by a 1000-run Monte-Carlo simulation. The solid line shows a Gaussian fit. The time delay corresponding to the peak of the Gaussian fit determines the spectral lag.} 

\label{fig:ccf}
\end{figure}
\begin{figure}

\begin{center}
\resizebox{3.4in}{!}{\includegraphics{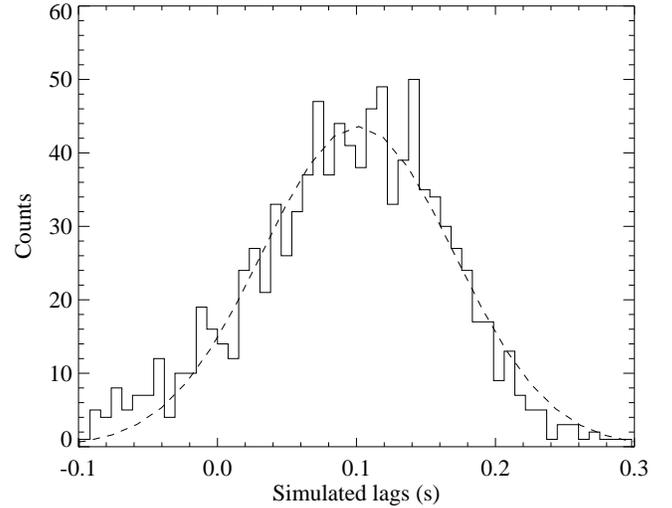}}
\end{center}
\caption{Histogram of 1000 simulated spectral lag values for the second sub-burst between 25-50 keV and 50-100 keV in Episode II. The dashed line shows a Gaussian fit. The standard deviation of the distribution of simulated spectral lag values is taken as its uncertainty.}
\label{fig:ccfsim}
\end{figure}
\begin{figure}

\begin{center}
\resizebox{3.4in}{!}{\includegraphics{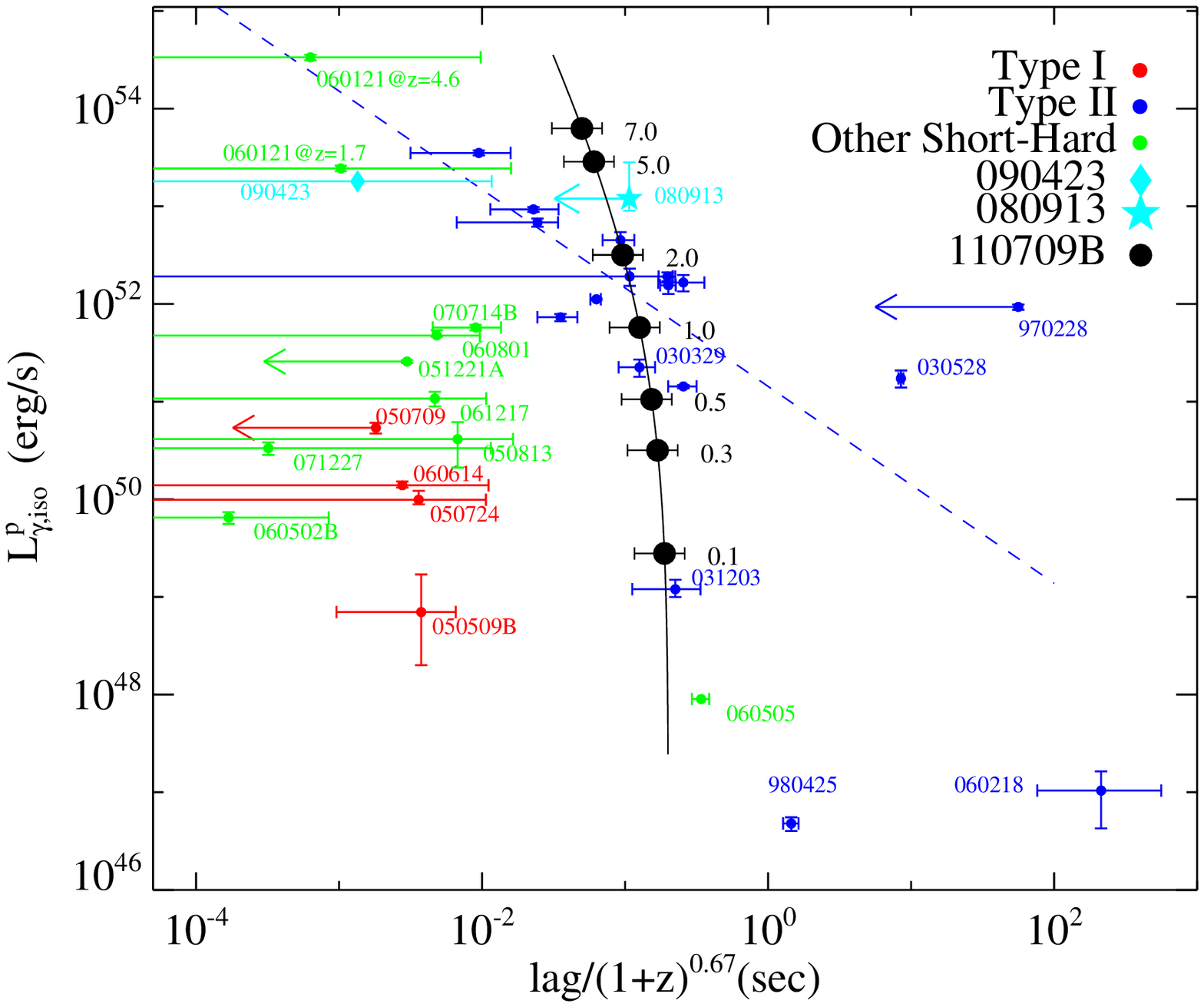}}
\end{center}

\caption{Luminosity-spectral lag diagram. Black points and solid lines indicate the double burst by putting it at different assumed redshfit $z_x$. Background blue data points are type II GRBs in Zhang et al. 2009. Background red and green data points are type I and ``other short-hard bursts" in Zhang et al. 2009. The dashed blue line represents the best linear fitting to type II bursts.}
\label{fig:lplag}
\end{figure}

Spectral lags, which are caused by the fact that softer Gamma-ray photons usually arrive later than hard photons, are always significant in long (type II) GRBs (Norris et al. 2000; Gehrels et al. 2006; Liang et al. 2006; Zhang et al. 2009), but are typically negligible for short (type I) GRBs (Norris \& Bonnell 2006; Zhang et al. 2009). In order to get high signal-to-noise ratio, we only select the brightest part (Episode I and II, as shown in Fig. \ref{fig:multilc} and listed in Table 2) of each sub-burst to calculate lags. For the first sub-burst, we extracted 32ms-binned light curves in the following four BAT energy bands: $15-25$ keV, $25-50$ keV, $50-100$ keV and $100-150$ keV and 64ms-binned light curves in the following three Konus-{\em{WIND}} bands: $25-95$ keV, $95-380$ keV and $380-1435$ keV. Then, using the CCF (cross-correlation function; Norris et al. 2000, Ukwatta et al. 2010) method, we calculate the lags between any two light curves in the neighboring and next-to-neighbor energy bands within each instrument in Episode I. The uncertainty of lags are estimated by Monte Carlo simulation (see e.g., Peterson et al. 1998, Ukwatta et al. 2010) and are illustrated in Fig. \ref{fig:ccf} and \ref{fig:ccfsim}. For the second sub-burst, we extracted 32ms-binned light curves in the same four BAT energy bands as mentioned above and three XRT energy bands: $0.3-1$ keV, $1-4$ keV and $4-10$ keV. Then using the same method we calculated the spectral lags between these energy bands. Our results are shown in Table 2. Some lags are not well-constrained possibly due to low signal-to-noise levels and the combination of multiple pulses. Yet the typical values of $201\pm52$ ms between $25-50$ keV and $50-100$ keV for the first sub-burst is similar with other long (type II) GRBs (Zhang et al. 2009). In Fig \ref{fig:lplag}, we plot the luminosity-lag diagram by assuming the double burst is at redshift $0.1-7.0$. We found that at the average redshift ($z\sim 2$) of {\em Swift} GRBs, GRB 110709B falls into the ``main track" of typical long/Type II GRBs in the luminosity-lag diagram.



\begin{table}

 
 \label{table2}
 
 \caption{Results of the spectral lag analysis} 



	   


 \begin{tabular}{lll|ll}
 \hline
 \hline

\multicolumn{3}{c|}{Energy Channels}   &   \multicolumn{2}{c}{Lag }       \\

& & &  \multicolumn{2}{c}{ (ms)}       \\
\hline

\multirow{2}{*}{Inst.}& $E_{low}$& $E_{high}$  &  Episode I  & Episode II \\
	   
& keV & keV & (0s,45s) &  (640s, 660s) \\
\hline
XRT &  0.3-2  & 2-4   & \multicolumn{1}{c}{-} & $682\pm 281$ \\
XRT & 2-4  & 4-10    & \multicolumn{1}{c}{-}  &  $ 208\pm 362$\\
XRT & 0.3-2  & 4-10    & \multicolumn{1}{c}{-}  & $ 1191\pm 339$\\
BAT & 15-25  & 25-50  & $118\pm 307 $ & $ 17 \pm 652 $   \\
BAT & 25-50 & 50-100  & $201\pm 52  $ & $107\pm  77$ \\
BAT & 50-100  & 100-150   & $ < 512 $ &$ < 3465$  \\
BAT & 15-25  & 50-100    & $277\pm 161 $ & $189\pm 232$  \\
BAT & 25-50  & 100-150  & $  <825 $ &  $ < 1741$  \\
KW  & 25-95  & 95-380  &$101\pm 148$  &  \multicolumn{1}{c}{-}  \\
KW  & 95-380  & 380-1435  &$< 1283$ & \multicolumn{1}{c}{-} \\
KW &  25-95  & 380-1435   & $< 2119$ & \multicolumn{1}{c}{-} \\

\hline \hline
\end{tabular}


\end{table}

\subsection{A Dark Burst ?}\label{sec:dark}

\begin{figure}
\begin{center}
\resizebox{3.4in}{!}{\includegraphics{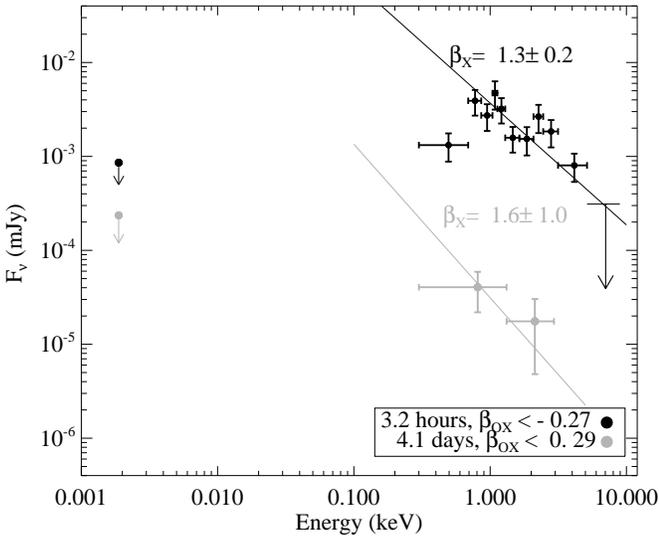}}
\end{center}

\caption{Optical to XRT band spectral energy distribution (SED) at $t=3.2$ hour and $t=$ 4.1 day. The R-band upper limits are obtained from Fong \& Berger 2011. Here spectral index $\beta$ is defined by $F_\nu \propto \nu^{-\beta}$. $\beta$ is calculated between the optical upper limits and peak X-ray flux. Solid lines are the power-law components fitted to XRT data only.}
\label{fig:optsed}
\end{figure}

There is no optical counterpart or host galaxy observed by UVOT or any other ground telescopes for GRB 110709B. Furthermore, no cataloged extragalactic galaxy was found within $1'$ radius in the NASA/IPAC Extragalactic Database (NED). Using the optical afterglow upper limits reported by Fong \& Berger (2011), we plot the optical-to-X-ray SED at $t=$ 3.2 hours and $t=$4.1 days (t is relative to trigger time $T_0$) in Fig. \ref{fig:optsed}. The corresponding $\beta_{OX}$ (spectral index $\beta$ is defined by $F_\nu \propto \nu^{-\beta}$.) are $< - 0.27$ and $< 0.29 $ for the two epochs. Since bursts with $\beta_{OX} < 0.5$ are defined as ``dark" (Jakobsson et al. 2004, van der Horst et al. 2009, Greiner et al. 2011), GRB 110709B is clearly an unusual dark burst with an even negative $\beta_{OX}$ (at $t=3.2$ hours). Furthermore, the EVLA detection of the radio counterpart of GRB 110709B gives further support that GRB 110709B is a dark burst (Zauderer \& Berger 2011).  With a large extragalactic soft X-ray absorption (\S 2.1), the absence of the optical afterglow detection probably indicates a very dusty ISM environment of GRB 110709B so its optical afterglow is highly extinguished. Alternatively, it may also indicate a high redshift origin (Fong \& Berger 2011) or very different radiation mechanisms between the X-ray and optical components (D'Elia \& Stratta, 2011). 


\section{Implications for the Central Engine }\label{sec:theory}

Long-term central engine activities have been proved by the commonly detected X-ray flares which occur at hundreds of seconds after the burst trigger. This double burst GRB 110709B suggests that the long-term active central engine not only powers X-ray flares but also can power a second burst. Generally speaking, in order to produce a second ``burst" as is observed in GRB 110709B, the central engine must restart with comparable or even larger energy. This is challenging for the following popular theoretical X-ray flare models:

\begin{itemize}
\item {\bf Fragmentation in the massive
star envelope.} The collapse of a rapidly rotating stellar core leads to fragmentation (King et al. 2005). If the delayed accretion of fragmented debris leads a second burst, the debris must have comparable masses with the materials in the initial major accretion. This behavior has not been seen to date in numerical simulations (e.g., Masada et al. 2007; Lee et al. 2009; Metzger et al. 2008; Lee et al. 2009).

\item {\bf Fragmentation in the accretion
disk.} Fragmentation of an accretion disk and subsequent accretion of the fragmented blobs may power X-ray flares in both short and long GRBs (Perna et al. 2006). In order to power a second burst instead of X-ray flares, the fragmented out part of the disk should have a comparable mass to that of the inner part of the disk, which is difficult to achieve. This model also predicts that later accretion (accretion of a blob farther away from the black hole) tends to spread in a longer duration, which is suitable to interpret X-ray flares, but not the double burst.

\item {\bf Magnetic barrier around the accretor.} Proga \& Zhang (2006) 
argued that a magnetic barrier near the black hole may act as an 
effective modulator of the accretion flow. The delayed outflow can power the X-ray flares. It is difficult for this model to account for the extreme energetics of the second sub-burst, since it is expected that a magnetic barrier can only block a smaller accretion rate, and hence, can only power a less violent episode such as X-ray flares.

\end{itemize}

On the other hand, the long quiescent gap between the two sub-bursts
leads us to re-think the 2-stage fallback collapsar scenario that has been used to interpret GRB precursors (Wang \& M{\'e}sz{\'a}ros, 2007). 
In that scenario, the precursor is produced by a weak jet
formed during the initial core collapse, possibly related to MHD processes associated with a short-lived proto-neutron
star, while the main burst is produced by a stronger jet fed by fallback accretion onto the black hole resulting from the
collapse of the neutron star.
We found that the assumed proto-neutron star rotational energy of a few times $10^{51}$ ergs in Wang \& M{\'e}sz{\'a}ros, 2007 would
also be sufficient, when beaming is taken into account, to power the first 
sub-burst of GRB 110709B. In fact, simple estimates indicate that
maximally rotating proto-neutron stars could reach rotational energies
as high as several $10^{52}$ erg. Here,
we propose a magnetar-to-BH scenario as follows:

(1) A magnetar is formed and produces the first sub-burst by
releasing its rotation energy via electromagnetic and
gravitational radiation in $\sim 10-20$ seconds (rest frame). 
A magnetar, rather than a lower field neutron star, is required not
only to produce the high luminosity ($L_{\gamma,iso} \sim 10^{52}\ {\rm erg}\
{\rm s}^{-1}$) and $E_{p,rest}$ ($\sim$ 0.6-1 MeV) of the first sub-burst
(Zhang \& M{\'e}sz{\'a}ros, 2001; Metzger et al. 2011), but also to
overcome the ram pressure of the fallback matter (Wang \&
M{\'e}sz{\'a}ros, 2007). For a typical magnetar with proto-neutron star radius $R_{PNS}\sim
50 \ {\rm km}$ and mass $M_0\sim 1.4 M_\sun$, the ram pressure can be
written as $P_{\rm ram}=\frac{\dot{M} v_{ff}}{4\pi R_{PNS}^2}\simeq
5\times10^{26}\dot{M}_{-2}M_0^{1/2}\left(\frac{R_{PNS}}{50{\rm
km}}\right)^{-5/2} {\rm erg\ s^{-1}}$, where $v_{ff}=(2GM/R_{PNS})^{1/2}$
is the free-fall velocity and $\dot{M}$ is the mass infalling rate in units of
$M_{\sun} {\rm\ s^{-1}}$ . The magnetic field pressure can be written
as $P_B=B_f^2/8\pi\simeq 4\times10^{28}B_{f,15}^2{\rm erg\ 
{\rm s}^{-1}}$. Comparing the two, one \footnote{Generally speaking, a relatively weaker magnetic field or a relatively longer initial rotation period leads to a longer magnetar spin-down time scale, and hence, the emission duration (Zhang \& M\'esz\'aros, 2001). For comparison, to interpret the long plateau ($\sim 16$ ks) in the X-ray light curve of GRB 070110, the magnetic field of a millisecond-period magnetar needs to be $B_f \sim  3\times 10^{14}\ G$ (Troja et al. 2007). For the case of GRB 060218 the initial spin-down period should be longer (e.g., $\sim 10 $ ms instead of $\sim 1$ ms) due to the low GRB energy constraint (Soderberg et al. 2006; Mazzali et al. 2006; Toma et al. 2007). See Lyons et al. (2010), Rowlinson et al. (2010) and Fan et al. (2011) for more individual examples. 
	}
can get $B_f \gtrsim 10^{14}\ G$. Such a magnetized jet internally dissipates and powers the observed gamma-ray emission (e.g. Zhang \& Yan 2011; Metzger et al. 2011).

(2) After the magnetar slows down, the magnetic outflow stresses decrease,
so the ram pressure of the infalling matter becomes dominant. Thus the
activity of the magnetar is suppressed during the accretion process. The accretion onto the magnetar does not lead to GRB emission, since the hot NS likely launches a dirty neutrino-driven wind with heavy baryon loading. In order to form a BH,
a total accreting mass of $1\ M_\sun$ is needed. Assuming a redshift z=2, the accretion rate is about $\dot{M}\sim
\frac{1M_\sun}{500s/(1+z)} \simeq 0.006 M_\sun/s$, which is consistent
with theoretical predictions in the supernova fallback scenario (see e.g., MacFadyen et al. 2001).

(3) The accretion finally leads the magnetar to collapse to a black hole. The second sub-burst is produced either from a baryonic or a
magnetic jet. The spectrum will be softer either because the
accretion leads the gas near the central engine to be more
baryon-loaded so that the jet is slower or because the pre-existing
channel from the first sub-burst may not have time to close so that
the wide channel results in a slower jet and a softer spectrum.
The spectral evolution of the two stages would be expected to be different,
since they are due to different physical process. These model features
appear to be in concordance with the observed facts (see Fig 
\ref{fig:ept}).

\section{Discussion}

\subsection{A Lensed Burst?}

\begin{figure*}
\begin{center}
\resizebox{7.0in}{!}{\includegraphics{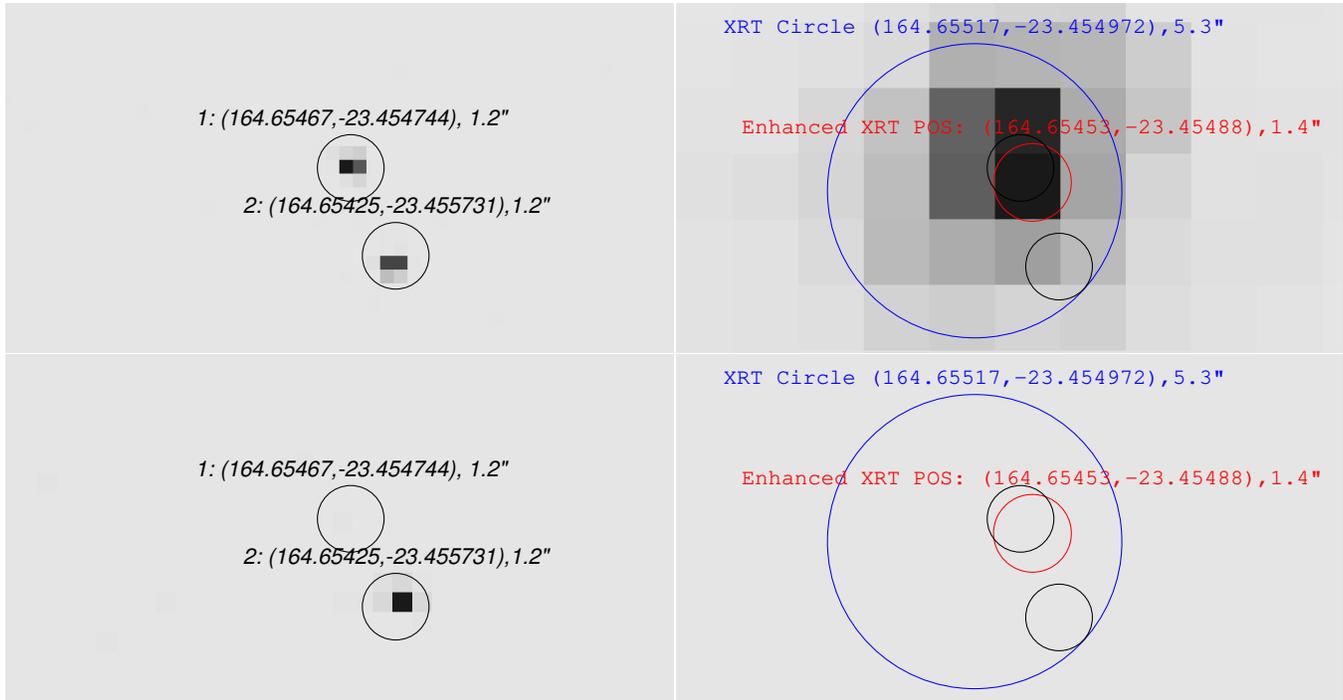}}
\end{center}
\caption{{\em Chandra} ({\it upper left}: $T_0 + 14\ days $; {\it lower left}: $T_0 + 114\ days$ ) and {\em Swift}/XRT ({\it upper right}: $T_0+0\ day$; {\it lower right}: $T_0 + 114\ days$) images of 110709B. Black circles (radius=$1\farcs 2$) indicate the Chandra source extraction regions at the locations of ${\rm R.A. (J2000)}=10^h58^m37.121^s$, ${\rm Dec. (J2000)}=-23^{\circ}27'17\farcs 08$ and ${\rm R.A. (J2000)}=10^h58^m37.003^s$, ${\rm Dec. (J2000)}=-23^{\circ}27'20\farcs 24$. The red circle is the enhanced XRT error circle (Beardmore et al. 2011) of the afterglow. The blue circle indicates the preliminary XRT error circle (based on the on-board centroid of the first 2.5 s of data) that was reported by Cummings et al. 2011.}
\label{fig:chandra}
\end{figure*}

The similarity of the two sub-bursts raises the question of whether they could be produced by gravitational lensing of a single GRB located behind a foreground galaxy. To investigate this possibility, we first examined Chandra observations of GRB 110709B at 14:15:04 UT on 23 July 2011 (day 14; 15.05 ks exposure time; Observation ID 12921) and at 19:50:34 UT on 31 October 2011 (day 114; 10 ks exposure time; Observation ID 14237). We downloaded the public Chandra data from the Chandra archive\footnote{\url{http://cda.harvard.edu/chaser}} and processed them using the standard CIAO tools (version 4.3). The first Chandra observation has two X-ray point sources in the field of GRB 110709B, with nearly identical brightness ($3.7\times 10^{-3}$ $ {\rm s}^{-1} $, 0.2-8 keV) and separated by only 3.4 arcseconds (Fig. \ref{fig:chandra}). Source 1 is located 0.67 arcseconds from the refined XRT position, within the refined XRT error circle. Both sources are within the XRT point-spread function (18 arcseconds Half-Power-Diameter), and the sum of their fluxes is consistent with the total XRT flux measured during the first epoch, while the flux of Source 1 is consistent with the extrapolation of the XRT light curve (Fig. \ref{fig:xrtlc}). The field was unobservable by both Chandra and Swift from about 8 August 2011 until 28 October 2011. In the second Chandra observation, taken shortly after the field emerged from the Chandra Sun (pitch angle) constraint, Source 1 has vanished, while Source 2 is still present, with a slightly lower count rate of $\sim 2.7 \times 10^{-3}$~s$^{-1}$ (0.2-8~keV), consistent with being a background X-ray source such as an AGN. The upper limit for the Source 1 flux is still consistent with the extrapolation of the XRT light curve (Fig. \ref{fig:xrtlc}). The fact that Source 1 vanished while Source 2 did not clearly rules out any possibility that the two Chandra X-ray sources in the double burst field are due to gravitational lensing. 

On the other hand, assuming the time delay ($\sim$ 11 minutes) between two sub-bursts is caused by gravitational lensing, we calculated that the angular separation of the two lensed images would be $\sim 10^{-2} $ arcsecond (Walker \& Lewis 2003), which is beyond Chanda best resolution capacity ($\sim$ 0.5 arcsecond). We found that a typical dwarf galaxy at $z\sim 1$ would be able to serve as the massive lensing object and cause such a separation. In this scenario, the difference between the pulse structure of the two sub-bursts can be understood by taking into account a structured jet and the so-called nanolensing effect (Walker \& Lewis 2003). However the different $E_p$ and spectral evolution (see \S 3) of the two sub-bursts are still difficult to explain. We thus disfavor the gravitational lensing explanation for this burst.

\subsection{A huge precursor, a huge X-ray flare, or a long quiescent gap?}

GRB 110709B is a unique event. It is one out of 613 GRBs detected by {\em Swift}/XRT so far (as of 30 December 2011; Evans et al. 2009, 2011). Since nearly half of {\em Swift} GRBs have X-ray flares (Maxham \& Zhang, 2009), it is roughly one out of $\sim 300$ GRBs with flares.  

Comparing with other GRBs, one may wonder whether this is a burst with a huge precursor (the first sub-burst), a huge X-ray flare (the second sub-burst), or simply a long GRB that has an extremely long quiescent gap in between. We address these possibilities in turn. 

\begin{itemize}
\item {\bf A huge precursor?} A good fraction\footnote{Observationally, this fraction is highly dependent on the definition of precursor and may suffer from instrumental bias. For example, Koshut et al. (1995) search a BATSE (Burst Alert and Transient Source Experiment) GRB sample and found the fraction is $\sim 3\%$. On the other hand, by using a different definition, Lazzati (2005) analyzed a sample of bright, long BATSE GRBs and found the fraction is $\sim 20\%$.} of GRBs have a precursor leading the main burst. A precursor is generally defined as an emission episode whose peak intensity is much lower than that of the main episode, and with a quiescent separation period from the main episode (Koshut et al. 1995; Burlon et al. 2008, 2009; Troja et al. 2010). Precursors may or may not trigger the Gamma-ray detectors (Lazzati 2005). Moreover, the peak energy ($E_p$ of the $\nu F_\nu$ spectrum) of the precursors is almost always softer than the emission. Some good examples of GRBs with typical precursors are GRBs 041219A (G{\"o}tz et al. 2011), 050820A (Cenko et al. 2006) and 060124\footnote{We note that the pulse structures of the main emission phase ($t>400$ s) of GRB 060124 and the second sub-burst of GRB 110709B are quite similar, namely a short duration pulse followed by the main emission, then an extra soft X-ray flare. See Fig. 4 in Romano et al. 2006 for comparison.} (Romano et al. 2006b). By contrast, the first sub-burst of GRB 110709B has a comparable intensity and harder $E_p$ than the second sub-burst, which is very different from a precursor. Nonetheless, some precursor models (e.g., Wang \& M{\'e}sz{\'a}ros, 2007) may give hints to the theoretical explanation of the double emission episode behavior of GRB 110709B (see \S \ref{sec:theory}).

\item {\bf A giant X-ray flare?} As discussed in \S1, X-ray flares are generally thought to be related to late time central engine activities. The shapes of X-ray flares are always soft in spectrum and smooth ($\delta t /t \ge 1$) in time profile (Burrows et al. 2005a; Chincarini et al. 2007; Falcone et al. 2007). In contrast, the X-ray emission from the second sub-burst of GRB 110709B shows a spiky time profile and a higher $E_p$ (up to $\sim 100$ keV) than those of a typical X-ray flare. Its X-ray fluence is comparable ($\sim$ 50\%) to the BAT fluence of the first sub-burst. The only giant X-ray flare that reaches a fluence comparable to prompt emission was GRB 050502B (Burrows et al. 2005a; Falcone et al. 2006). However, the flare of GRB 050502B was much softer, and smoother. We thus regard the X-ray emission from the second sub-burst is more analogous to prompt emission. Using the popular X-ray flare model to interpret the second sub-burst is challenging as discussed in \S\ref{sec:theory}.

\item {\bf A long quiescent gap?} GRB 110709B has a very long quiescent gap ($\sim 500$ s). We note that this gap is unusual but not unprecedented. For example, significant long quiescent periods have been observed in some other long GRBs, such as GRB 070721B ($t_{gap}\sim 200$ s ; Ziaeepour et al. 2007) and GRB 091024B ($t_{gap}\sim 500 $ s; Gruber et al. 2011). On the other hand, GRB 110709B is unique in the sense that the two sub-bursts separated by the gap have somewhat similar pulse shapes, comparable emission durations, comparable peak intensities, and comparable fluences. We thus regard GRB 110709B as a unique {\it double} burst. Nevertheless, there are still some similarities between it and other bursts with long gaps (especially GRB 091024, which has three comparable emission episodes). The model we propose in \S\ref{sec:theory} might be applicable to those bursts as well.

\end{itemize}

\section {Summary}

GRB 110709B is the first GRB with two {\em Swift}/BAT triggers. We present in this paper a comprehensive study on Gamma-ray and X-ray observations of this unusual GRB and its afterglow. No optical afterglow or host galaxy has been detected for this burst. By putting this burst at redshift $\sim 2$ (average redshift of {\em Swift} GRBs), we found it can be a typical long (Type II) bursts which follows previous empirical relations (such as Amata relation, lag-luminosity relation) quite well. The dark burst nature of GRB 110709B may suggest a very dusty environment or high redshift origin or different radiation mechanisms between X-ray and optical band. Although separated by 11 minutes, the two sub-bursts may physically originate from the same central engine, which apparently requires extreme two-step activities that may be related to magnetar-to-BH accretion. 


\begin{acknowledgements}

We thank John Nousek, Judith L. Racusin, Alexander J Van Der Horst, Peter Veres, Hao-Ning He, Mark Walker, Davide Burlon, David Gruber, Eveline Helder, Jonathan Gelbord, Zach Prieskorn, En-Wei Liang, Suk Yee Yong and Fuwen Zhang for helpful comments. This work is partially supported by the following grants: NASA SAO SV4-74018 [BBZ], NASA NNX08AL40G [PM], NSF PHY-0757155 [PM], NSF AST-0908362 [BZ], NASA NNX10AD48G [BZ], ASI grants I/009/10/0 [GS], NSFC 10973008 [XYW] and the 973 program 2009CB824800 [XYW]. The Konus-Wind experiment is supported by the Russian Space Agency and the Russian Foundation for Basic Research (grants 09-02-00166 and 11-02-12082).

\end{acknowledgements}

\end{document}